# Evaluation of frictional suspension system performance in three-piece bogie by the multi-body-system approach

Reza Serajian*, Mehrnoosh Abedi, Iman Hazrati Ashtiani
University of California Riverside, CONCAVE Research Center, Concordia University, Montreal, QC, Canada
* Corresponding author's email: rsera004@ucr.edu

## Abstract

The nonlinear damping characteristics of friction wedges in the secondary suspension of a freight bogie are investigated considering unidirectional contact and non-smooth frictional forces. In the proposed MultiBody System model, each wedge has six Degrees of Freedom with corresponding inertial properties. The geometry of wedge, as well as, wedge angle, toe-in condition, and clearances between wedge/bolster and wedge/side frame are also considered in the modeling. The methodology for the identification of contact parameters is presented to achieve a smooth response and efficient numerical solutions. The response of the system with the proposed method is compared with the quasi-static methodology of wedge simulation and limitations of alternative methods are highlighted. The model is then applied to evaluate the effects of different design parameters of the wedge system on vertical and lateral response of the system.

# 1 Introduction

The early studies of railway vehicle lateral stability were based on linearized models of wheel/rail contact and suspension components. However, all kinematic and constitutive relations in wheel-rail contact, as well as truck and suspension components, are inherently or intentionally nonlinear. The highly nonlinear frictional suspension is common in freight railway cars due to low maintenance and production cost but dynamic modeling and numerical treatment of them have been yet a challenging issue. The value of linear analysis for a complex system such as a three-piece railway truck cannot be underestimated as it provides highly valuable tool during preliminary studies and design of component parameters, geometries and profiles. On the other hand, realistic prediction of highly complex response and dynamic behavior of such system is only possible through detail nonlinear simulation and analysis. From the literature review it is evident that even in most complex and detail nonlinear analysis simplifications are used to represent the friction damping present in the secondary suspension and the sophisticated ones have not been used in complete vehicle dynamics simulations. The design introduces set of complex friction wedges for variable multi-directional damping in the secondary suspension to enhance the performance limits over a wide range of loading condition.

The present investigation considers modeling of friction wedges in three-piece freight bogies. Three-piece bogie consists of two side frames, sets on the axles, and a central beam called bolster which is connected to side frames by two spring nests and wedges. Damping in this suspension design is provided by frictional elements in wedges, center plate and side bearings. The general concept of this design has been relatively unchanged over the years but several generations of this bogie have been proposed by modification in elements as well as side bearings and wedges to provide safer operation and higher productivity.

Different combination of wedges and suspension system in three-piece bogies can be classified as variable and constant frictional damping, shown in Figure 1. The generated normal force in interaction faces of wedges is related to deflection of wedge and bolster springs. On the other hand, in constant damping design presented in Figure 1(b), the preloaded spring of wedge generates a constant normal force in wedge-side frame and wedge-bolster contact faces. The generated frictional forces are also related to friction coefficients between wedge-side frame, $\mu_s$, and wedge-bolster, $\mu_b$. Toe condition of the side frame, shown in Figure 1(a), is another design parameter that can be divided as no-toe ($\gamma=0$), toe-out ($\gamma>0$) and toe-in ($\gamma<0$). Toe angles have implications not only for the movements of components but also for forces within the wedge suspensions.

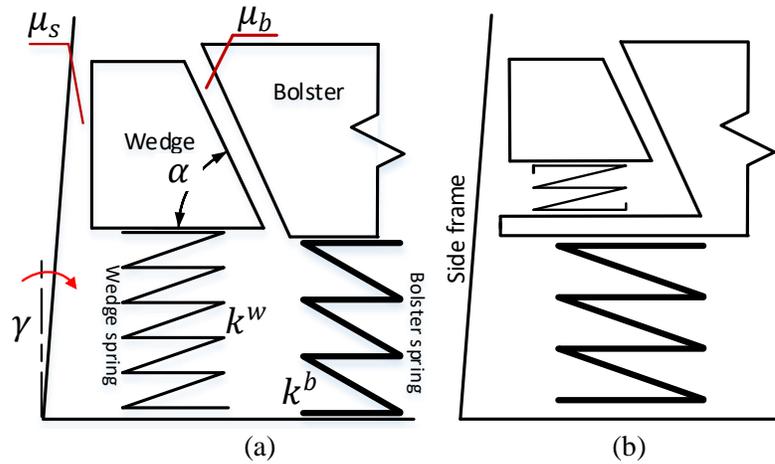

Figure 1. Different configurations of wedges in three piece-bogies, (a) variable damping and (b) constant damping.

Various modeling approaches have been applied to capture some of the mechanical and geometrical constraints of this bogie. Accurate predictions of the critical speeds and dynamic responses, however, necessitate in-depth nonlinear component modeling and analysis. An extensive review of different approaches for the modeling of wedges in three-piece bogie is presented by Wu *et al.* [1]. Wu *et al.* [2] used a "white-box" suspension model focusing on original wedge suspensions with three different toe angles and proposed a methodology which could be applied to optimize wedge suspensions. The different angle parameters of wedges in three-piece bogies are investigated and substantially improved the response of the system, especially in warp (lozenge) stiffness using improved friction wedge designs by Tournay [3]. Iwnicki *et al.* [4] also reviewed the development of three-piece freight bogies and introduced innovative modifications in components.

The hunting stability of railway vehicles was studied in different studies with a focus on different parameters to obtain the sensitivity of the system with respect to variation of suspension parameters [5, 6] and wedge angle [7, 8]. The nature of the contact forces and dissipation of excitation through friction forces are also investigated in the response of the system in longitudinal train dynamics studies [9, 10]. The friction wedges in the secondary suspension of a three-piece truck exhibit strongly nonlinear damping property, which is attributed to complex variations in the contact forces, contacting surfaces geometry and the friction coefficient [11, 12]. Moreover, the friction wedge design yields variable multi-directional damping of the secondary suspension to enhance the performance limits over a wide range of loading conditions. Xia [13] considered the effects of wedge mass and introduced the damping to the system by two-dimensional dry friction model. Dry friction generates the stick-slip condition between two moving parts and poses a variable state space formulation of the system. When the stick condition is presented, the wedge and bolster are considered as a single body. For such condition, the equations of motion of the system are derived as follows based on free body diagram of the wedge and bolster presented in Figure. 2.

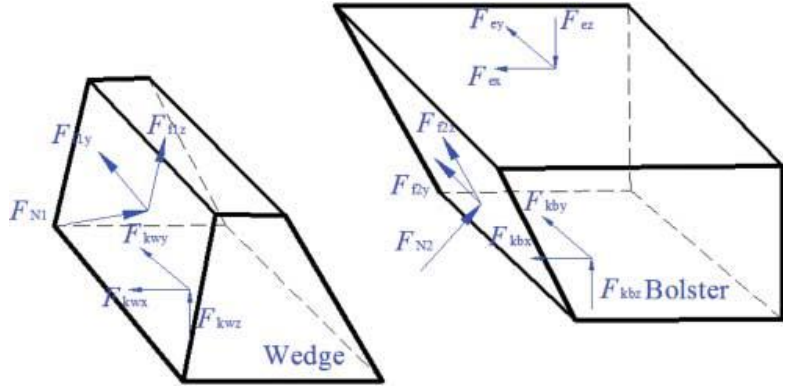

Figure 2. Two-dimensional friction wedges model presented by Xia [13].

$$m_w \ddot{x}_w = (F_{N1} \cos(\gamma) + F_{f1z} \sin(\gamma)) \times \Gamma + (F_{f2z} \cos(\alpha) - F_{N2} \cos(\alpha)) \times \Psi - F_{kwx}$$
$$m_w \ddot{y}_w = (-F_{f1y}) \times \Gamma + F_{f2y} \times \Psi - F_{kwy} \quad (1)$$
$$m_w \ddot{z}_w = (F_{N1} \sin(\gamma) - F_{f1z} \cos(\gamma)) \times \Gamma + (F_{N2} \sin(\alpha) + F_{f2z} \cos(\alpha)) \times \Psi - F_{kwz} + G_W$$

$$m_b \ddot{x}_b = (F_{N2} \cos(\alpha) - F_{f2z}) \times \Psi - F_{ex} - F_{kbx}$$
$$m_b \ddot{y}_b = -F_{f2y} \times \Psi - F_{kby} - F_{ey} \quad (2)$$
$$m_b \ddot{z}_b = (-F_{f2z} \cos(\alpha) - F_{N2} \sin(\alpha)) \times \Psi + F_{ez} - F_{kbz} + G_b$$

Where $x$, $y$, $z$ (downward positive) are displacement of bolster and wedge; $w$ and $b$ indexes correspond to wedge and bolster; The normal contact forces, $F_{Ni}$ are generated in wedge/side and wedge/bolster faces are indicated by $i=1$ and 2 indexes. Wedge and toe- angles are $\alpha$ and $\gamma$; gravity components of wedge and bolster are considered in $z$ directions as $G_w$ and $G_b$ and reaction forces of coil springs are $F_{k..}$ in $x$, $y$ and $z$ directions for wedge and bolster. The interface forces of bolster and carbody are also presented as $F_e$ in formulations; the switch conditions of equations of motions are considered as $\Gamma$ and $\Psi$.

Subject to consideration of pure slip condition of friction forces, $\Gamma$ and $\Psi$ parameters are 1 and the kinetic constraints of the system could define as:

$$\dot{x}_w = -\frac{\sin(\gamma)\sin(\alpha)}{\sin(\gamma+\alpha)} \dot{z}_b, \quad \dot{z}_w = \frac{\cos(\gamma)\sin(\alpha)}{\sin(\gamma+\alpha)} \dot{z}_b \text{ and } \dot{x}_b = 0 \quad (3)$$

The separations of bolster and wedges are considered when the excitation forces are large enough to generate $z_b - z_w > 0$ condition. The longitudinal kinematic constrain of bolster, $\dot{x}_b$, is not negligible anymore and the kinematic constrains in Eq. 3 are no longer valid and the $\Psi$ parameter must switch to 0 for Eq. 2 for separated condition. In this condition the bolster perform forced vibrations without friction forces. The state of wedge after separation from bolster fall into two conditions. While the wedge spring is still loaded, there will be a contact between wedge and side frame, then only the switch condition, $\Psi$ in Eq. 1 is zero. Moreover, the longitudinal acceleration of the wedge should assumed as:

$$\ddot{x}_w = \ddot{z}_w \tan(\gamma) \qquad (4)$$

When the displacement of wedge is larger than initial value, wedge will have force vibrations under wedge springs loads and all contact forces are zero by switched the $\Psi$ and $\Gamma$ values to zero.

The stick condition of the friction forces causes a high-frequency variation in small relative velocities of components and thus demands a significant increase in the integration steps for a successful simulation. Xia [12], thus applied switching conditions to treat friction forces and proposed the structure varying system. The results showed that for small amplitude of excitation the motions of wedge and bolster are coupled and that friction plays an important role to prevent vertical resonance of the bolster. It was also observed that lateral excitation of the bolster will cause both lateral and vertical vibration in the wedges.

Harder [11] developed a specific element in *ADAMS* software for a given type of wedge presented in Figure 1(a). Ballew *et al.* [14, 15] investigated the effects of geometry and inertial properties of wedges on dynamic responses of a four-degrees-of-freedom (DOF)-half-truck model considering motions in the vertical, lateral, pitch and yaw directions. The study indicated that the magnitudes of friction forces obtained from the model were considerably lower than those from the *NUCARS* simulations. Steets [16] developed a three-dimensional (3-D) multi-body dynamic model for characterizing friction wedge interactions with the bolster and the side frame and discussed the key differences in responses when compared with the *NUCARS* simulation results.

Kaiser *et al.* [16] investigated dynamic responses of friction wedges assuming stick-slip Coulomb friction using a single-DOF model. The response to low amplitude and low-frequency vertical excitations was dominated by the sticking phase of the contact. The response to larger amplitude or higher frequency excitations was weakly nonlinear due to dominant slippage phase of the contact. The experiments conducted on a scaled model of a wedge confirmed the two types of slip-stick conditions and revealed occurrence of slip phenomenon at frequencies above 30 *Hz*.

Modeling of friction wedge suspension can be classified into three types: Combination models [e.g., 11, 12], quasi-static models [e.g., 13, 17] and Multi-Body System models (*MBS*) [16]. The advantage of combination models are the computation efficiency and simplicity of model but pose limitations to parameter analysis of wedge suspension design. Quasi-static models could evaluate the geometric parameters of wedge in addition to wedge-bolster separation. The *MBS* models, on another hand, are able to reveal the dynamic phenomena of system that combination or quasi-static models are not able to present as well as curved surfaces of wedge or warp stiffness of system. Moreover, application of penetration contact and friction contact models are only possible in *MBS* models. Despite this, Wu *et al.* [1] are indicated that *MBS* wedge models are not yet suitable for stability analysis in complex train system due to complexity and computation efficiency.

In this study, the secondary suspension system of a three-piece bogie is modeled in detail with multidirectional effects of spring nests as well as the friction wedge design. The detail model developed in Universal Mechanism (*UM*) software [18] further included the backlashes in suspension along with nonlinearity in suspension stiffness and friction. In addition to a detail discussion on identification of parameters for proposed method of friction wedges, a further objective of the simulation of suspension system of this bogie is to establish a quantitative study of the effects of different parameters of wedges on directional response of system for different excitations. Since the study examines the effects of suspension friction wedge design, the following section presents a detail consideration of this element.

## 2 Multi-body model of system

Depending on the purpose of investigation, there can be various levels of simplifications in the modeling of suspension elements. Among them, the shape, size and configuration of the wedges in secondary suspension are typically ignored and simplified by representing their effects as vertical or lateral equivalent dampers. Although the wedge mass is small compared to other components, it is designed to provide multi-dimensional friction forces and to be a function of loading on the bolster. In a realistic simulation, the effect of wedge shape and friction levels in the lateral, vertical as well as against yaw and warp motions under different loading would play an important role in the lateral dynamics and hunting behavior of a three-piece freight truck. In the following section, the methodology for simulation of different components of suspension in three-piece bogie is presented.

### 2.1 Coil Spring Modelling

The helical springs are generally considered as linear springs in the axial direction [1, 7]. In the secondary suspension of the bogie, each wedge is seated on a set of helical spring, while the bolster is seated on a five groups of spring sets. The couplings between different motions and the reaction moments, however, cause transverse and bending deformations of the springs leading to forces and moments along the non-axial directions [19]. The effective stiffness of the suspension springs is thus defined considering the coupling among translational and rotational motion and non-axial reaction forces such as:

$$\begin{bmatrix} k_{xx} & 0 & 0 & 0 & k_{xx}H/2 & 0 \\ 0 & k_{yy} & 0 & -k_{yy}H/2 & 0 & 0 \\ 0 & 0 & k_{zz} & 0 & 0 & 0 \\ 0 & -k_{xx}H/2 & 0 & k_\theta & 0 & 0 \\ k_{yy}H/2 & 0 & 0 & 0 & k_\beta & 0 \\ 0 & 0 & 0 & 0 & 0 & k_\psi \end{bmatrix} \begin{bmatrix} x \\ y \\ z \\ \theta \\ \beta \\ \psi \end{bmatrix} = \begin{bmatrix} F_x \\ F_y \\ F_z \\ M_x \\ M_y \\ M_z \end{bmatrix} \qquad (5)$$

In the above matrix, $k_{xx}$, $k_{yy}$ and $k_{zz}$ are the effective stiffness of the wedges and bolster springs. $k_\theta$ and $k_\beta$ are the bending stiffness of the spring corresponding to angular deformations about *x*- and *y*-axis ($\theta$ and $\beta$), respectively. $k_\psi$ is torsional stiffness attributed to deformation $\psi$ about the *z*-axis. *x*, *y* and *z* are the longitudinal, lateral and vertical directions, respectively. The effective stiffness of the spring along each axis are related to the spring geometry, material properties, actual height and number of coils of the inner and outer springs and given by [19].

$$k_{zz} = \frac{Gd^4}{64nR^3}; \quad k_\psi = \frac{Ed^4}{128nR}; \quad k_{xx} = k_{yy} = \frac{1}{\frac{1}{2(1+\nu)k_{zz}} + \frac{H^2(2+\nu)}{24k_\psi}}$$

$$k_\theta = k_\beta = \frac{2k_\psi}{(2+\nu) - \frac{H^2(2+\nu)}{8k_\psi\left(\frac{1}{2(1+\nu)k_{zz}} + \frac{H^2(2+\nu)}{6k_\psi}\right)}}$$

(6)

where, $E$ and $G$ are Young's and shear moduli of the spring material, $\nu$ is the Poisson's ratio. $d$ is the wire's diameter, $H$ is spring height, $R$ is the coil radius and $n$ is number of active turns. Table 1 summarizes the equivalent stiffness values of the wedge and bolster springs, which were computed for typical spring designs of bogie ($d$=20 $mm$, $R$=56 $mm$, $n$=6.45 for inner spring; $d$= 30 $mm$, $R$=85 $mm$, $n$=4 for the outer spring). While $k_{zz}$ and $k_\psi$ are independent of spring height, other components of stiffness are sensitive to variations of $H$. The height of the springs was taken as 256 $mm$, which the effects of variations in the height was neglected together with the damping due to springs. The effective stiffness of wedge and bolster, $k^w$ and $k^b$ are presented in Table 1.

Table 1. Effective stiffness coefficients of the secondary suspension supporting the wedge and the bolster

| Component | $k_{zz}$ (kN/m) | $k_{xx} = k_{yy}$ (kN/m) | $k_\theta = k_\beta$ (kNm/rad) | $k_\psi$ (kNm/rad) |
|---|---|---|---|---|
| Wedge spring, $k^w$ | 520 | 445 | 11 | 4.5 |
| Bolster spring nest, $k^b$ | 2,930 | 2,397 | 73/97* | 25 |

*The bending stiffness of the bolster spring nest due to the longitudinal/lateral deformations of the spring.

## 2.2 Frictional Wedges Modelling

The point-plane contact model available in *UM* platform [20] is used to estimate the normal and friction forces in each contact points. The general concept of the normal contact force is shown in Figure 3. Modeling normal forces in point-plane contact generally assumes as an elastic-dissipative function of penetration Δ in the contact point *A*, as:

$$F_N = K_c \Delta + C_c \dot{\Delta} \tag{7}$$

where $K_c$ and $C_c$ are the contact stiffness and damping constants, respectively. The direction of $F_N$ is align with contact plane normal vector, $\vec{n}$.

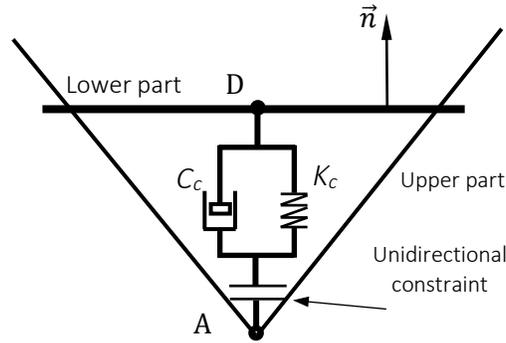

Figure 3. Schematic of the point-plane contact model.

Generated normal contact force of $F_N$, is a unidirectional force and applied in positive penetrations. The metal to metal contact case is the general condition of contact in the suspension system of a freight railway car. High values of contact stiffness in this condition lead to stiff system of differential equations which not only imposes greater computational demand but also contribute to noise in the system response [1, 20].

The tangential friction force of the contact is calculated based on Coulomb friction model and realizes the stick-slip condition. Sliding friction forces is evaluated considering Coulomb's friction and Newton's law in the opposite direction of sliding as $-\mu_s F_N \vec{v}_s/|\vec{v}_s|$, where $\mu_s$ is the dynamic friction coefficient and $\vec{v}_s$ is the sliding velocity vector. The dynamic friction coefficient $\mu_s$ is defined as function of the sliding velocity considering the Stribeck velocity $v_{str}$, such that [20]:

$$\mu_s(v_s) = \mu_\infty + (\mu_0 - \mu_\infty)e^{-\left(v_s/v_{str}\right)^\delta} \tag{8}$$

where $\mu_\infty$ is friction coefficient at a high sliding velocity and $\delta$ is an empirical exponent [20]. The Stribeck effect is based on the experimental observations that the drop in static friction coefficient to sliding one is not discontinues and has a characteristics as illustrated in Figure 4. On the other hand, the slip mode is simulated when $|F_s| > \mu_0 N$ where $\mu_0$ is static friction coefficient.

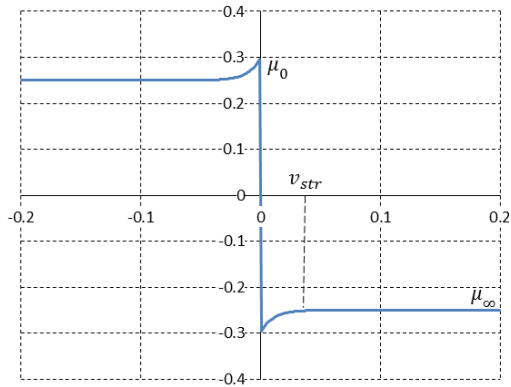

Figure 4. Stribeck effect in friction coefficient [20].

This formulation of contact has been used to simulate the force response and geometric constraint of the contacting pairs.

The friction damping in the truck elements is modeled considering the wedge geometry and its coupling with the bolster and the side frame. Each wedge in the truck suspension is modeled as a 6-DoF dynamic system considering its inertial properties. The domain of the wedge contact surface with the side frame is described by the coordinates of four corner points in the lateral plane of the wedge and the side frame, $A_1..A_4$, as shown in Figure 5. The contact domain of the wedge and the bolster is defined in a similar manner through coordinates of wedge edges and the bolster, $B_1..B_4$.

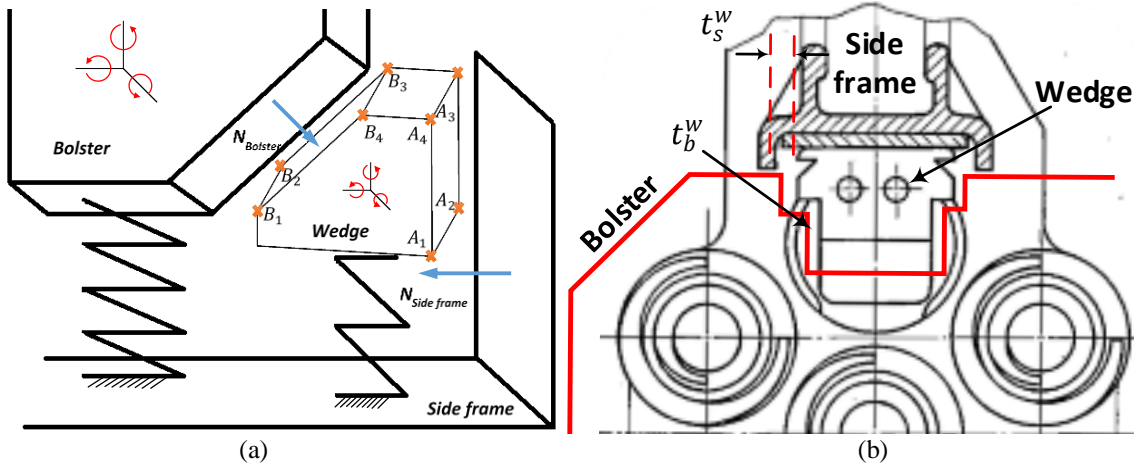

(a)　　　　　　　　　　　　　　　　　　　　　(b)
Figure 5. (a) Schematic of the contact points-planes between the wedge and the bolster/side frame, (b) clearances between components.

The relative lateral movement between wedge and bolster, $t_b^w$, as well as wedge and side frame, $t_s^w$, are shown in Figure 5(b), are limited by defined contact elements with a gap. These clearances can increase due to wear and plastic deformations and have a significant effects on resultant warping and lozenging performance of bogie. As initial setup, $t_b^w$ and $t_s^w$ are defined as 10 and 50 *mm*.

The contact stiffness of each contact element is computed assuming a high natural frequency compare to dominant frequency range of dynamic response, which is known to occur only up to 20 *Hz* [1, 20, 21]. The adjustment of contact parameters, especially $K_c$ and $C_c$, require considerations in terms of computational demand and force response of system. The initial value of contact stiffness of each contact point of wedge is initially estimated assuming the contact partial frequency of 100 *Hz* and mass of the lighter body, namely wedges in suspension system of bogie. Response of a benchmark system subject to similar input will then evaluated for high contact stiffness. The adequate stiff contact could be selected where variation of dynamic responses of system subject to stiffer contact properties becomes negligible.

A similar approach is also used for identification of contact damping. The correct choice of damping is very important especially for unilateral contacts and modelling of gap or impact. The value of contact's damping coefficient is assumed to be correlated with contact's stiffness and notion of damping ratio, $\zeta$. The initial values of contact damping for each pair of contact are calculated as: $C_c = 2\zeta\sqrt{mK_c}$.

## 2.3 Modelling of virtual test rig

The dynamic characteristics of the suspension wedges under different loading conditions are established using a virtual test rig of the bogie created in the *Universal Mechanism* software. The test rig isolates the effects of wheel/rail and track clearances and consists of two side frames, bolster of bogie and four wedges. Table 2 present the inertial and some geometric parameters of model including the mass centers (*c.g.*), $I_x$, $I_y$ and $I_z$ that are mass moments of inertia of components about *x*-, *y*- and *z*-axis of the body-fixed coordinate system with origins located at the *c.g.* of the respective component.

Table 2. Inertial and geometric properties of the wagon components [8, 13, 20, 22].

| Component | Mass (kg) | $I_x$ (kg×m²) | $I_y$ (kg×m²) | $I_z$ (kg×m²) | c.g. height* (m) | Additional Geometric parameters |
|---|---|---|---|---|---|---|
| Bolster | 596 | 323 | 7.25 | 326.3 | 0.701 | |
| Side-frame | 526.3 | 13.6 | 175 | 161.8 | 0.480 | y** = ± 0.978 m |
| Wedge | 21.6 | 0.08 | 0.103 | 0.102 | 0.567 | x*** = ± 0.335; y** = ± 0.978 m |

* *c.g.* heights with respect to the top of the rail; ** lateral position with respect to the middle of the track *** longitudinal position with respect to the center plate.

The effects of side bearings and center plate in addition to the wagon inertia are neglected in simulated system. The side frames are fixed to the ground and a virtual actuator is used to move the bolster in a defined direction. For instance, the given harmonic displacement of bolster, $Z_{bolster}$ and $y_{bolster}$ in vertical and lateral directions are defined as:

$$Z_{bolster} = A_z(sin(2\pi f_z t)) + \delta_{initial} \quad ; \quad y_{bolster} = A_y(sin(2\pi f_y t)) \tag{9}$$

where $\delta_{initial}$ is the initial deflection of bolster (represent payload of wagon), $f_z$ and $f_y$ are frequency of the vertical and lateral oscillations in *Hz*, and $A_z$ and $A_y$ are the magnitude of directional displacement. Two different values for $\delta_{initial}$ define the loaded and unloaded condition of bogie. The given displacement of the bolster generates the reaction forces in different directions which are the resultant forces of spring deflection, inertial forces and contact forces developed in different interfaces of contacting parts. The forces on the secondary suspension could also be varied by assigning a vertical preload on the bolster to simulate loaded and unloaded condition of wagon.

## 2.4 Identified contact parameters of wedges

The proposed virtual test rig could thus be conveniently used to examine the overall responses of the wedges for varying contact stiffness and damping in a highly efficient manner. This proposed setup permits the observation of the stick-slip condition and behavior of the contact elements under excitation of different amplitudes and frequencies. The vertical excitations of bolster in unloaded condition are used to identify the contact parameters of wedges. The resonance un-damped frequency of bolster could be assumed as 5.47 *Hz* for unloaded condition base on approximate [23]:

$$f_z = 1/2\pi \times \sqrt{2k_{zz}^b + 4k_{zz}^w/m_b + 0.5 \times m_w} \tag{10}$$

where $k_{zz}^b$ and $k_{zz}^w$ are stiffness of bolster and wedges springs; $m_b$ and $m_w$ are, masses of bolster and wagon, respectively. The carbody is assumed to be 12 400 *kg* in unloaded case as a typical condition [7, 22]. Based on discussions in section 2.2, the initial contact stiffness of $K_c$=8.29 *MN/m* is assumed to obtain the contact frequency of 100 *Hz* as a simplified single DoF system. Moreover, the static and sliding friction coefficient, $\mu_0$ and $\mu_\infty$ respectively, are assumed 0.30 and 0.25 for wedges-side frame contact surfaces while 0.12 and 0.10 are assumed as friction coefficients in wedges-bolster contact surface [1, 22]. The position of contact points is defined based on geometry of wedge while the wedge angle is defined as 45° and non-toed condition. The contact stiffness is varied from the initially assumed value of 8.29 to 165.8 *MN/m*, and the responses are obtained considering two different damping ratios (0.1 and 0.2) due to contact damping, $C_c$. The equations of motion for the virtual test rig are solved using the Park method [24, 25] with time step varying from $1\times10^{-12}$ to a maximum of 0.001 *s*, and error tolerance of $1\times10^{-8}$.

The consequences of stiffer contact stiffness and higher damping ratio of contact are presented in Figure 6 for unloaded condition of bolster; $\delta_{initial} = 7\ mm$, $A_z = 5\ mm$ and excitation frequency of 5.47 *Hz*. The graphs have a clockwise rotation while positive values of *x*-axis present the loading cycle of excitation.

In displacement range of ±5 *mm* the bolster reaches to extremum position with small velocities where sharp changes in contact forces generate. As presented in Figure 6, the base values (8.29 *MN/m*) of contact stiffness leads to a partly soft system with notable oscillations. These low-frequency oscillations tend to convergence to a general trend of force-displacement loops after 5 times stiffer contact elements, shown in Figure 6(a). On another hand, the integration time step was generally around 0.1 *ms* for simulations corresponding to baseline contact stiffness, which gradually decreased to 0.025 when the contact stiffness was increased 20 times (169 *MN/m*), suggesting substantially higher computational demand with higher contact stiffness.

As discussed before, low damping ratio and high stiffness parameters of contact elements cause the high-frequency oscillations of response. This trend is clearly shown comparing Figures 6(a and b), where higher damping ratio of $\zeta = 0.2$ presents a lower noise amplitude. The results thus suggest the convergence of the solutions with contact stiffness more than 58 *MN/m*, irrespective of the contact damping values considered in the simulation.

The contact stiffness equal to 7 times the base value of 8.29 *MN/m* together with 0.1 contact damping ratio was subsequently chosen as a compromise to achieve non-oscillatory and convergent solutions. Table 3 summarizes the wedge-bolster and wedge-side frame contact parameters, which are subsequently used for further analyses.

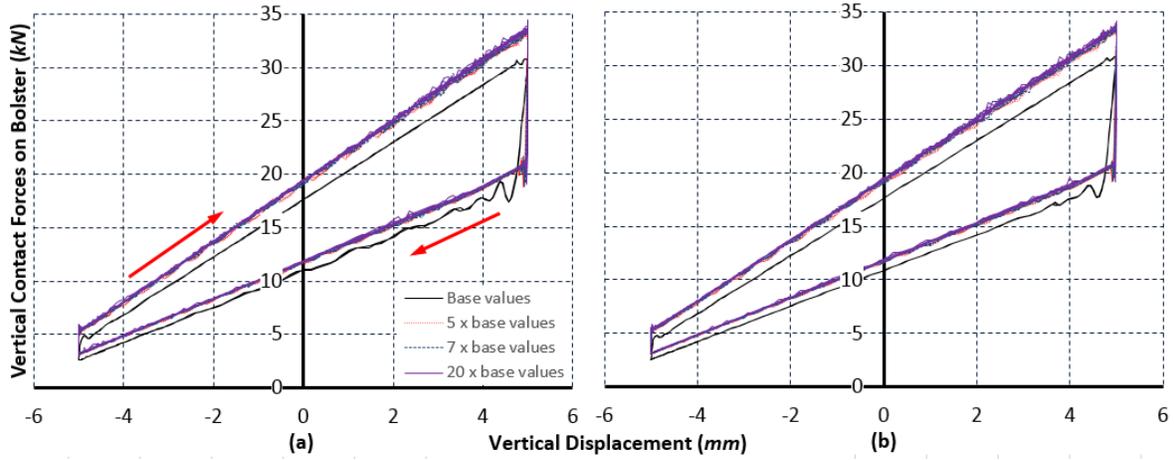

Figure 6. Effect of variation of contact parameters for (a) $\zeta = 0.1$ and (b) $\zeta = 0.2$.

Table 3. Stiffness, damping and friction coefficients of the contact pairs.

| Contact between | $K_c$ MN/m | $C_c$ kN×m/s | $\mu_0$ | $\mu_\infty$ | $\delta$ | $v_{str}$ mm/s |
|---|---|---|---|---|---|---|
| Wedge and bolster | 58 | 6.98 | 0.12 | 0.1 | 1 | 1 |
| Wedge and side frame | 58 | 6.98 | 0.30 | 0.25 | 1 | 1 |

In force-displacement graph, four different patterns could be identified. In lower part of the loop, the bolster spring nests are undergoes an unloading or extension course. The compression course of springs and higher normal forces in contact points generates a relatively higher equivalent stiffness of system comparing the extension cycle. The transfer from slip to stick while the sliding velocity is small and springs are compressed generates a higher jump comparing to extension cycle. The observed difference is due to higher vertical components of contact forces in compressed condition of springs. The general trend of force response is correlated with presented results of different studies as well as [14, 15, 23] although the exact matching of the response is not possible due to different properties and limited provided information.

The identified stiffness in loading and unloading cycles of excitation with proposed method can compare with proposed methodology of Xia [13] summarized in Eques 1 to 4 to establish a quantitative judgment. Considering quasi-static assumptions, pure vertical displacement and zero toe angle of side frame, the vertical component of applied contact forces on bolster, $F_c^b$, can be presented as follows:

$$F_c^b = k_{loading} \times z_w = 4 \times k_{zz}^w \left( \frac{1 + \mu_b \cot\alpha}{1 + (\mu_b - \mu_s)\cot\alpha + \mu_b \mu_s} \right) \times z_w \qquad (11)$$

for loading cycles and;

$$F_c^b = k_{unloading} \times z_w = 4 \times k_{zz}^w \left( \frac{1 - \mu_b \cot\alpha}{1 - (\mu_b - \mu_s)\cot\alpha + \mu_b \mu_s} \right) \times z_w \qquad (12)$$

for unloading cycles.

The equivalent stiffness of loading and unloading cycles, assuming the sliding coefficients presented in Table 3 are 2.61 and .59×10⁶ *N/m* respectively. On the other hand, the equivalent stiffness of compression and extension cycles are approximately identified as 2.79 and 1.77×10⁶ *N/m*. The proposed wedge model, comparing to quasi-static approach, presents a relatively higher equivalent stiffness which could be linked to more accurate simulation of stick-slip frictional force variations in addition to consideration of wedge inertial properties. Moreover, the proposed model is able to capture mechanical phenomenon, as well as multi-directional response of system which is not possible with other available methodologies

## 3. Results and Discussions

The test rig simulation results indicate the effectiveness of the methodology for validating or characterizing a complex element prior to its use in a dynamic simulation of the entire system. The proposed model is used to evaluate the effects of different parameters of friction wedges. The results are classified in vertical and lateral directions of excitation inputs in the following sections.

### 3.1 Effects of loading condition of wagon

The identified parameters of contact elements are used for evaluation of suspension system response under a free oscillations of loaded and unloaded conditions of wagon. Free oscillations of bolster from an initial position while the carbody and cargo are assumed as 12 400 and 58 000 *kg* to generate the loading condition of 200 *kN* on each wheelset, applicable for conventional design of bogie. The response is presented in Figure 7.

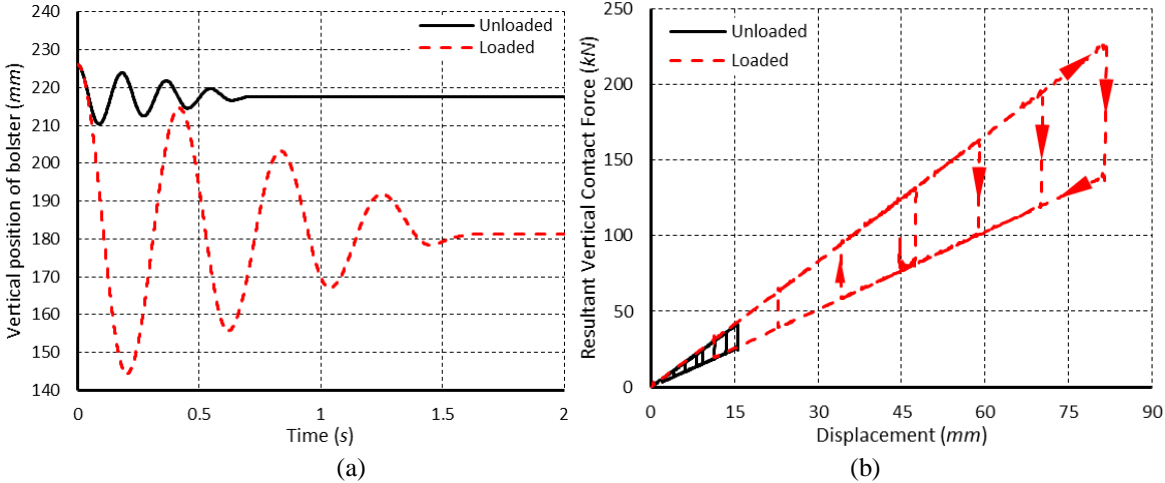

Figure 7. Free vertical oscillations of bolster under unloaded and loaded conditions, (a) variation of vertical position, (b) force vs displacement variations.

The frequency of oscillation in unloaded condition is identified as 5.42 *Hz* while in loaded condition of bogie the frequency is dropped to 2.43 *Hz*. On the other hand, the calculated damping ratio by logarithmic decrement shows the damping ratio of oscillation in unloaded condition as 0.3% while damping ratio in loaded condition is increased to 2%. The identified frequency of oscillation in this method shows the considerable effects of friction wedges on simple vertical response of system comparing to the simplified method such as Equation 10 especially in loaded conditions.

The new identified natural frequencies of system is used in test rig to evaluate sensitivity of proposed model on variation of excitation frequency. For loaded conditions of wagon, the $\delta_{initial}$ is adjusted as 30 *mm* to represent the pre-deflection of suspension in this loading condition. Hysteretic variation of force under frequency of 5.42 *Hz* and 2.43 *Hz* for unloaded and loaded condition and amplitude of 5 *mm* are illustrated in Figure 8. The response in unloaded conditions shows a negligible level of oscillation and robustness of proposed model with variation of excitation frequency.

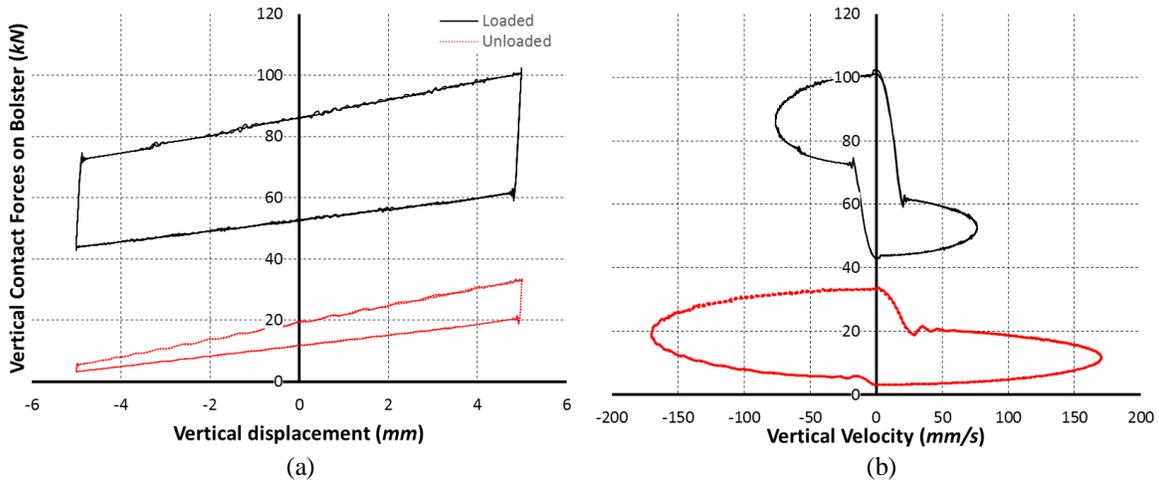

(a) (b)

Figure 8. Variation of contact force on bolster in loaded and unloaded condition (a) force-displacement and (b) force-velocity graphs.

The test rig is used for evaluating the performance of suspension system subject to lateral excitation of bolster in loaded and unloaded conditions. For lateral excitation $\delta_{initial}$ is defined as 7 and 30 *mm* for unloaded and loaded conditions while lateral amplitude and frequency of excitations, $A_y$ and $f_y$, are defined as 15 *mm* and 2 *Hz* based of Equation 9. The generated contact force on bolster in lateral direction are presented verse lateral displacement and it velocity in Figure 9. The clearances between bolster and wedges, defined as 10 *mm* for $t_b^w$ in Figure 5(b), have a major role in the pattern of force response. In unloaded condition, the clearance of ± 10 *mm* is clearly identified by negligible variations of contact forces in lateral direction and sliding of bolster over the wedge. The spikes in force response are due to collision of wedge and bolster in lateral plane. On the other hand, higher applied forces in loaded condition reduced the gap of 10 to 6 *mm*.

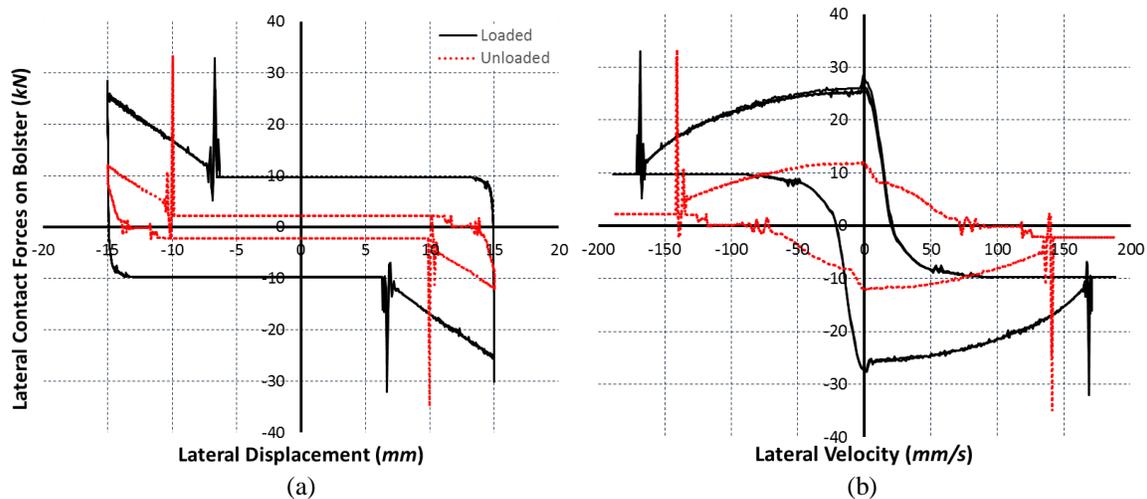
Figure 9. Variation of lateral component of contact forces on the bolster in loaded and unloaded condition: (a) force-displacement and (b) force-velocity.

It has been shown in Figures 8 and 9 that the generated damping in the system for vertical and lateral excitations are lower for unloaded condition which is in agreement with reported analyses in several studies [7, 26]. The stability analyses of the conventional three-piece bogies are usually performed in unloaded conditions because the lateral performance of suspension system is lower than loaded condition. Due to this fact and for the sake of brevity, the following studies on effects of different components of bogie are reported for unloaded condition in following sections.

## 3.2 Effects of wedge angle

Variation of vertical and lateral response of suspension system is investigated subject to variations of wedge angle in the parametric model. Figure 10 shows the effect of wedge angle variation between wedges and bolster on resultant friction forces due to vertical and lateral harmonic displacements. The higher wedge angle leads to higher oscillation in resultant friction forces and higher damping level of the system. Moreover, subject to lateral harmonic excitations of bolster in unloaded condition, the higher wedge angle generates higher frictional forces although the effects are not of much consideration.

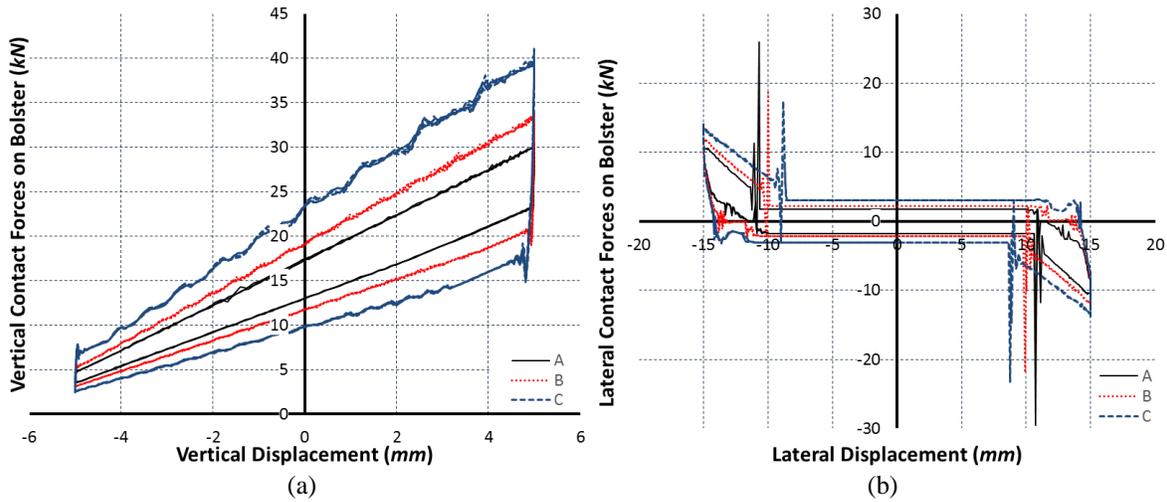

Figure 10. Effects of wedge angle; A=30°, B=45° and C=60°, variation on resultant friction forces on bolster for (a) vertical and (b) lateral excitation of unloaded condition.

## 3.3 Effects of toe-in configuration of the wedge

The toe angle of between wedge and side frame generates a coupling between lateral and vertical displacements of wedge as presented in Equations. 4 and 5. Toe-in geometry of 1 degree, increase the resultant vertical contact forces in loading and unloading cycles. On other hand, the 1° toe-out angle the variations are not considerable comparing non-toed condition but forces are slightly lower in loading and unloading cycle. This new pattern is correlated with provided longitudinal movement due to the toe-out configuration and higher contact forces.

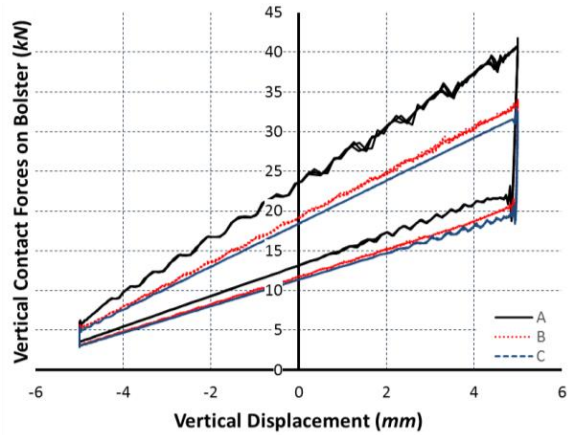

Figure 11. Effects of toe angle; A: 1° toe-in; B: zero-toe and C: 1° toe-out.

## 3.4 Effects of variations of friction coefficients

The performance of frictional wedges subject to variations of frictional coefficient between side frame and wedges are evaluated comparing the base values presented in Table 3 with 25 percent higher and lower values. The response of the system subject to the vertical and lateral excitations of the bolster is presented in Figure 12. The friction coefficients in wedges and bolster interfaces are assumed constant to simplify the analysis. The higher friction coefficient generates higher vertical forces in loading cycles while the resultant vertical forces are lower in unloading part of excitation. In lateral excitation, the range of ±10 *mm* has not been affected because of direct relation of this part to wedge-bolster interface. The effects variation in of friction coefficient in ± 15-10 *mm*, while the wedges are sliding over the side frame, are not as considerable as effects of wedge angle variations.

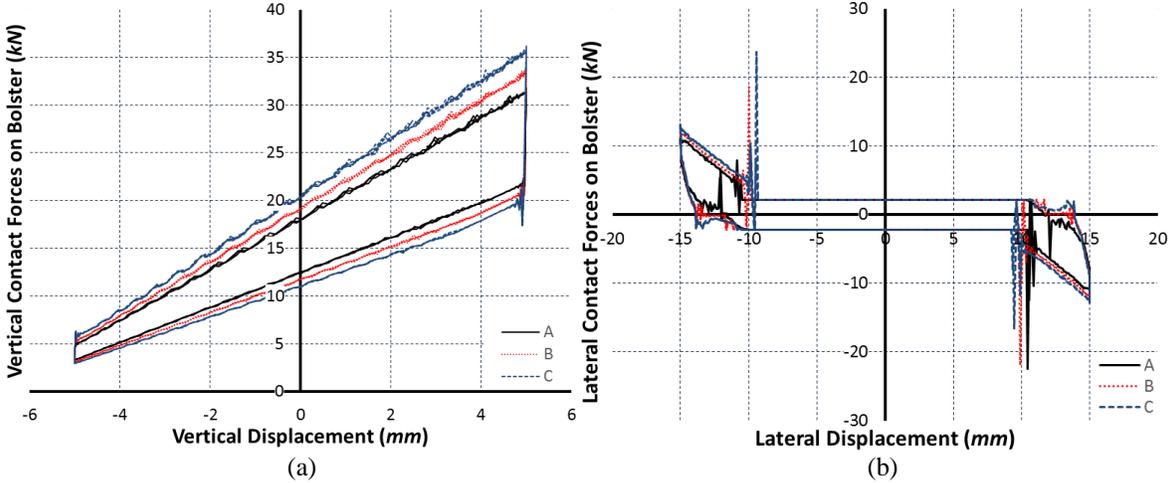

Figure 12. Effects of friction coefficient variation between wedges and side frame on resultant a) vertical and b) lateral forces of the system; A: 25 % lower, B: base and C: 25 % higher values of friction coefficient.

The friction coefficient of incline interface between wedge and bolster could change during the application of system or due to lubrication practice. Figure 13 illustrates the effects of modified wedge/bolster friction coefficient on resultant contact forces subject to vertical and lateral excitations. In vertical direction, increase of friction coefficient slightly changed the response by reducing the slope of loading cycle. In lateral direction, the major change is in margin of ±10 *mm* which is related to sliding of bolster over the wedge surface. The increase in friction coefficient increases the magnitude of resultant contact forces in the lateral direction. For larger lateral displacement between 10 and 15 *mm*, there is not any relative motion between bolster and wedge and variations friction coefficient in this interface do not have any effect on response.

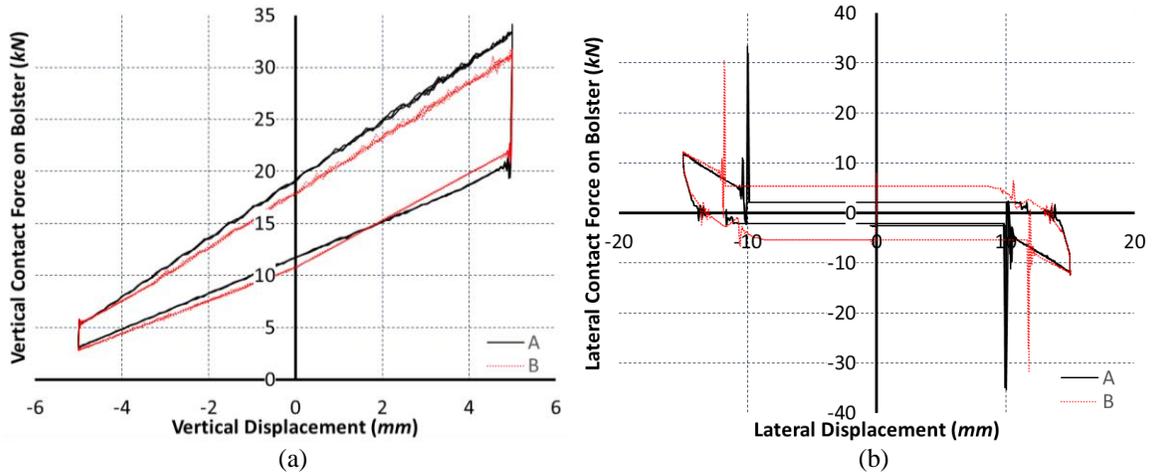

(a)                               (b)

Figure 13. Effects of friction coefficient variation between wedges and bolster on resultant (a) vertical and (b) lateral contact forces of bolster for A: 0.1 and B: 0.25.

## 3.5 Warp performance of suspension system

The warp performance of three-piece bogie is the ability of suspension system to react to lozenging inputs and stay in square shape. In general warp performance could be related to friction performance of axle boxes, center plate and side bearings in addition to the friction moments of wedges and warp stiffness of springs. To evaluate the warp performance of the model, the side frames of bogie undergoes a longitudinal 0.5 *Hz* harmonic movement with 100 *mm* amplitude while the left and right side frames are 180 degrees out phase. In proposed model, contribution of axle box and center plate are not simulated. The applied displacement inputs generate angular distortion of wedge components and bolster's rotation in yaw direction, as presented in Figure 14(a). The warp performance of suspension as variation of applied moment versus warp angle of bolster is illustrated in Figure 14(b) for unloaded and loaded conditions. The configurations of wedge geometry and contact parameters are as presented in Table 3. The unloaded condition has a lower warp restraint comparing to loaded condition. Simulated gaps and clearances between wedge-bolster and wedge-side frame have major effects of variations of warp stiffness in different loading conditions. These clearances could dramatically change due wear which generates a different response. The abrupt slope changes of response are due to these clearances, specifically in ± 50 *mrad* margin where wedge and bolster are in contact laterally. During the lateral movements of side frames, wedges experience planar and rotational displacements, since they are not positioned at the center of rotation of bolster. The movements generated frictional forces which generate extremely non-linear performance of stiffness and damping of system. The proposed *MBS* modeling of wedges can efficiently be applied for evaluation of different configurations and design parameters of system.

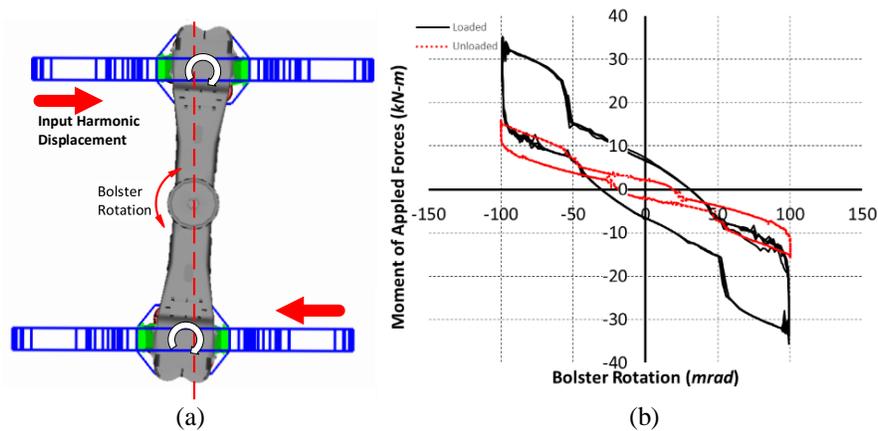

(a) (b)

Figure 14. The bogie warp test (a) schematic diagram of the test and (b) variations of applied moment with respect to yaw rotations of the bolster.

## 4. Conclusion

The focal point of this study was development a multibody dynamics model to simulate non-smoothness of friction forces in suspension system of three-piece bogie and friction wedges. Different components of three-piece bogie as rigid bodies are interconnected to each other with elastic coil springs and contact elements in *MBS* simulation package of Universal Mechanism (*UM*). A virtual test rig is developed in *UM* to characterize the effects of different wedge variables with respect to vertical and lateral movements of bogie. The stick-slip condition in addition to non-linear variation of stiffness and damping in compression and extension are also incorporated for lateral and vertical excitation. Different contact surfaces of the wedges are simulated as pairs of contact points-plane. Normal contact force in each contact point-planes is calculated based on a linear relationship between penetration and relative velocity. Non-smooth dynamic response of contact due to friction is simulated by considering static and dynamic friction coefficient in addition to Stribeck effect. The contact parameters are identified to achieve non-oscillatory and convergent solutions while the computation efficiency is maintained. The method provides the possibility of simulating the wedge geometric configuration including, inclination of contact surfaces, gapes and wear of friction wedges. Lower damping of the system is observed for un-loaded and lower friction coefficient for both lateral and vertical excitations when compared with loaded and high friction cases. Higher wedge inclination has a major effect on higher level of damping in the system. Moreover, lower friction coefficient between wedge and bolster has similar effect of damping increase. The development of the multibody dynamics model gives an opportunity to improve wedge modeling by creating more realistic models. Improved wedge modeling can more accurately predict detrimental freight train phenomenon that can lead to derailment such as wedge lock up and hunting.